\documentclass[letter]{jpsj3} 
\usepackage{misc}

\title{
Electric Polarization Induced by N\'{e}el Order 
without Magnetic Superlattice:
Experimental Study of Cu$_{3}$Mo$_{2}$O$_{9}$
and Numerical Study of a Small Spin Cluster}

\author{Haruhiko \textsc{Kuroe}\thanks{E-mail address: kuroe@sophia.ac.jp}, 
Tomohiro \textsc{Hosaka}, 
Suguru \textsc{Hachiuma},
Tomoyuki \textsc{Sekine}, 
Masashi \textsc{Hase}$^{1}$, 
Kunihiko \textsc{Oka}$^{2}$, 
Toshimitsu \textsc{Ito}$^{2}$,
Hiroshi \textsc{Eisaki}$^{2}$,
Masashi \textsc{Fujisawa}$^{3}$\thanks{Present address: Research Center for Low Temperature Physics, Tokyo Institute of Technology, Tokyo 152-8551 Japan},
Susumu \textsc{Okubo}$^{3}$, and 
Hitoshi \textsc{Ohta}$^{3}$
}%

\inst{
Department of Physics, Sophia University, 7-1 Kioi-cho, Chiyoda-ku, Tokyo 102-8554, Japan \\
$^{1}$National Institute for Materials Science (NIMS), Tsukuba, Ibaraki 305-0047, Japan \\
$^{2}$National Institute of Advanced Industrial Science and Technology (AIST), Tsukuba, Ibaraki 305-8568, Japan\\
$^{3}$Molecular Photoscience Research Center, Kobe University, 1-1 Rokko, 657-8501 Kobe, Japan\\
}

\abst{
We clarify that 
the antiferromagnetic order 
in the distorted tetrahedral quasi-one-dimensional spin system 
induces electric polarization.
In this system, 
the effects of low dimensionality and magnetic frustration 
are expected to appear simultaneously.
We obtain the magnetic-field-temperature phase diagram 
of Cu$_{3}$Mo$_{2}$O$_{9}$ by studying the dielectric constant 
and spontaneous electric polarization.
Around the tricritical point at 10 T and 8 K, 
the change in the direction of electric polarization causes 
a colossal magnetocapacitance.
We calculate 
the charge redistribution 
in a small spin cluster consisting of two magnetic tetrahedra 
to demonstrate 
the electric polarization induced by the antiferromagnetism.
}

\kword{multiferroic materials, frustrating magnet, 
charge redistribution, 
Mott insulator,
distorted tetrahedral spin system}

\begin{document}
\maketitle

\begin{figure}[b]
\begin{center}
\includegraphics[width=0.9\textwidth]{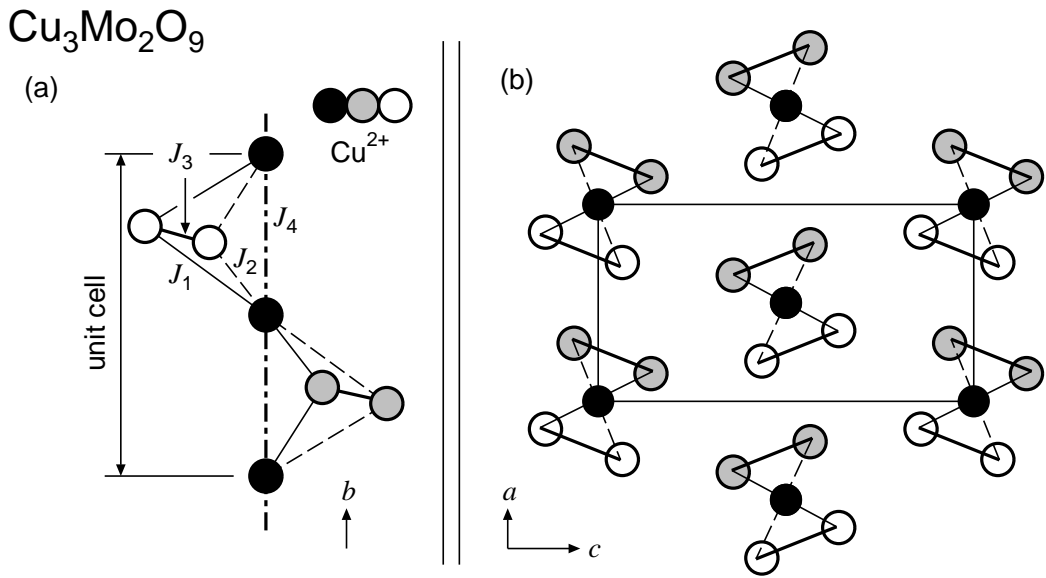}
\end{center}
\caption{
Schematics of the distorted tetrahedral chain 
in Cu$_{3}$Mo$_{2}$O$_{9}$ along the $b$-axis (a) 
and in the $ac$-plane (b). 
The circles indicate 
the $S$ = 1/2 Cu$^{2+}$ ions and 
the symbols distinguish their coordinates along the $b$-axis.
O$^{2-}$ and Mo$^{4+}$ ions are omitted.
The dashed, solid, bold, and dot-dashed lines 
distinguish the superexchange interactions $J_{1}$-$J_{4}$
between Cu$^{2+}$ ions.
The solid rectangle in (b) denotes the unit cell, 
which contains two tetrahedral chains. 
}
\label{TopView}
\end{figure}
Since the discovery of 
a strong magnetoelectric effect in TbMnO$_{3}$,~\cite{KimuraNature2003}
multiferroics in transition-metal oxides 
have been extensively studied.~\cite{CheongNature2007}
In the cases of the inverse Dzyaloshinskii-Moriya interaction 
in a spiral spin structure~\cite{Katsuura2005,Kenzelmann2005,Mostovoy2006}
and the inverse Kanamori-Goodenough interaction 
in a collinear one,~\cite{Arima2005} 
the formation of the magnetic superlattice 
plays an essential role in the 
magnetic-order-induced multiferroics.
Geometrical magnetic frustration 
also plays an important role 
as the origin of a nontrivial spin configuration 
that breaks the spatial inversion symmetry.
In this study, we demonstrate that 
the distorted tetrahedral spin system has 
the potential to show multiferroic behavior
{\it without} any magnetic superlattice formation.
We focus on the dielectric properties 
induced by an antiferromagnetic (AFM) spin order 
in Cu$_{3}$Mo$_{2}$O$_{9}$ 
and discuss the possibility of 
the multiferroic behavior in an AFM spin system 
on the basis of the theory of the charge redistribution 
in frustrated Mott insulators.~\cite{Bulaevskii2008,Khomskii2010}

Cu$_{3}$Mo$_{2}$O$_{9}$ has 
two distorted tetrahedral quasi-one-dimensional 
quantum spin systems 
of $S$ = 1/2 spins along the $b$-axis 
in its orthorhombic unit cell [Figs. \ref{TopView}(a) and 1(b)].
This compound has 
geometrical magnetic frustrations 
due to the tetrahedral spin alignment 
and 
quasi-one-dimensionality 
simultaneously.
This compound undergoes an AFM phase transition 
at $T_{\rm N}$ = 7.9 K without 
a magnetic field.~\cite{Hamasaki2008,Hamasaki2010} 
Inelastic neutron scattering measurements clarify 
the hybridization effects due to 
the $J_{1}$ and $J_{2}$ superexchange interactions 
between two elemental magnetic excitations, 
i.e., that of the quasi-one-dimensional AFM spin system 
originating from the $J_{4}$ (= 4.0 meV) superexchange interactions 
and that of the isolated AFM spin dimers 
originating from the $J_{3}$ (= 5.8 meV) 
ones.~\cite{KureJPCS,KuroeSubmittedToPRB}

The AFM spin-wave branch arises linearly 
from zero energy at 
the magnetic zone center of $(h, k, l)$ = (0, 1, 1), 
which is also the nuclear zone center 
below and above $T_{\rm N}$.
This magnetic branch is 
strongly dispersive along the $(0, k, 0)$ direction.
This dispersion curve and 
the two-spinon continuum
above mass gap energy $\Delta$ = 1.2 meV 
are 
typical features of a quasi-one-dimensional AFM spin system.
There are no indications of 
the formation of 
commensurate or incommensurate magnetic superstructures, 
at least at our experimental resolution.

\medskip
To obtain the magnetic-field-temperature ($H$-$T$) phase diagram, 
we measured the $T$ and $H$ dependences 
of the dielectric constant $\epsilon_{\alpha}$ 
and the electric polarization $P_{\alpha}$ 
under an electric field $E_{\alpha}$ 
along the $\alpha$-axis ($\alpha$ = $a$ or $c$) 
of Cu$_{3}$Mo$_{2}$O$_{9}$.
In this letter, we focus on 
the effects of the magnetic field 
along the $c$-axis.
We prepared a platelike single crystal of Cu$_{3}$Mo$_{2}$O$_{9}$
whose cross section and thickness 
are typically about 60 mm$^{2}$ and 0.4 mm, respectively.
To form a capacitor, 
the faces 
were coated with gold and attached using two gold wires.
The capacitance, 
of which the typical value was on the order of 10 pF, 
was measured using 
an impedance analyzer (Yokogawa-Hewlett-Packard 4192A).
$\epsilon_{\alpha}$ was obtained from the capacitance 
at 100 kHz with a peak voltage of 1 V.
To confirm the ferroelectric behavior, 
the electric polarization-electric field loop 
($P_{\alpha}$-$E_{\alpha}$ loop) 
at 1 Hz 
was recorded 
using a modified Sawyer-Tower circuit
with a peak voltage of about 200 V.
The magnetic field was applied using a superconducting magnet 
(Oxford Instruments, Teslatron S14/16), 
of which the maximum magnetic field was 16 T, 
and a variable-temperature insert cryostat was used to 
set the temperature from 1.5 to 300 K.

\medskip
\begin{figure}
\begin{center}
\includegraphics[width=\textwidth]{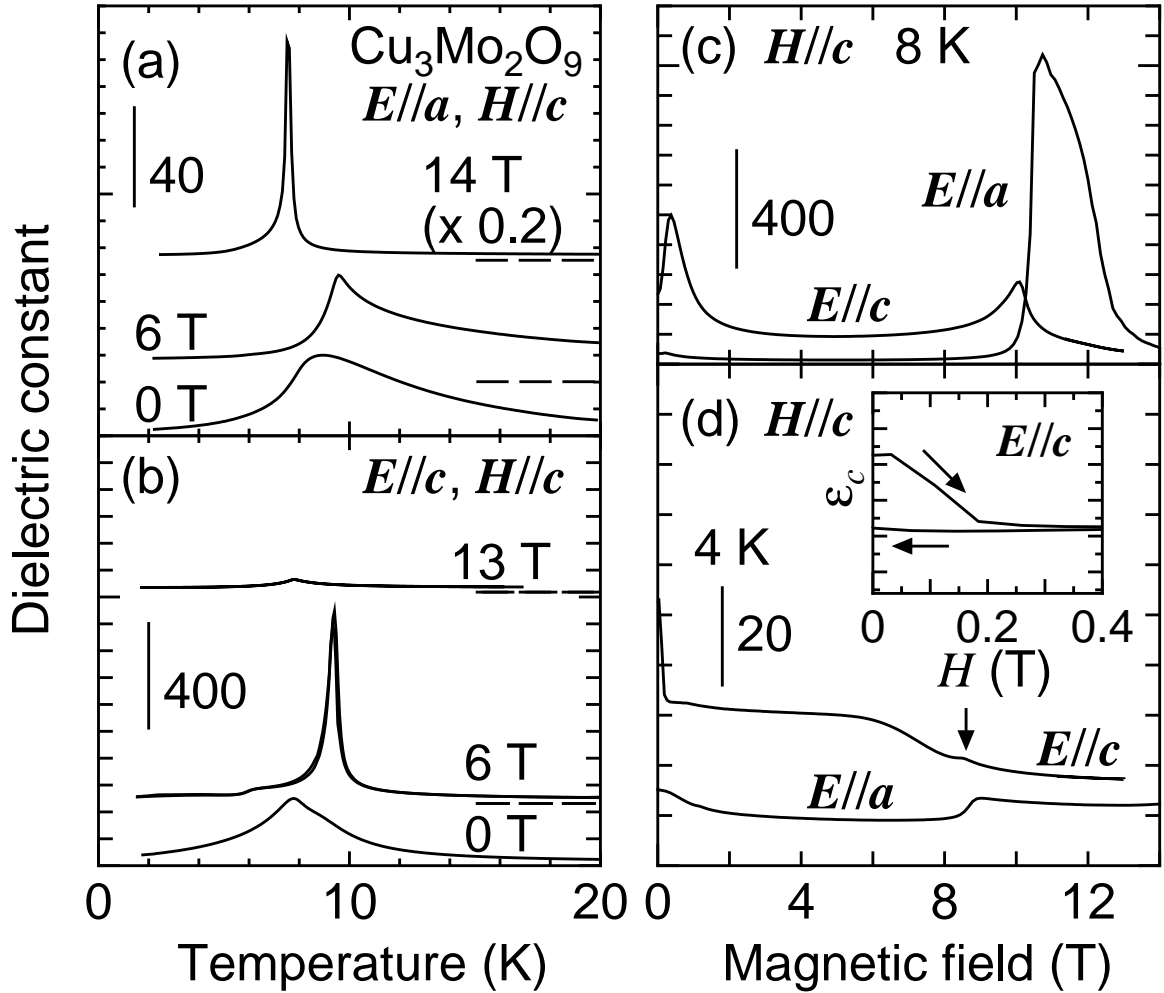}
\end{center}
\caption{
Typical temperature dependences of the dielectric constants 
under fixed magnetic fields [(a) and (b)].
For visibility, the data were shifted.
The magnetic field dependences of the dielectric constants 
at 8 and 4 K are shown in (c) and (d), respectively.
}
\label{DielectricConstantC}
\end{figure}
\begin{figure}
\begin{center}
\includegraphics[width=\textwidth]{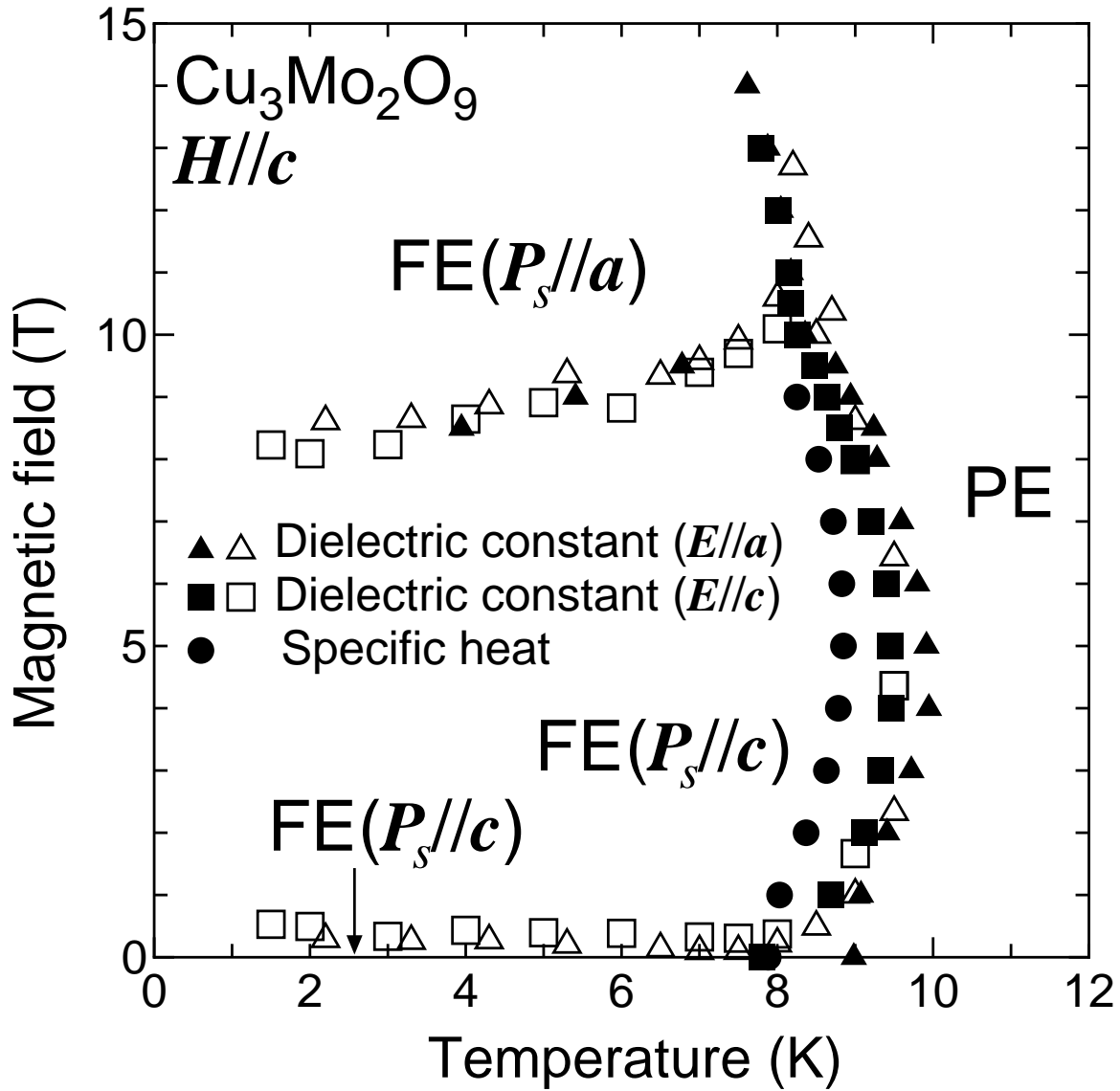}
\end{center}
\caption{
$H$-$T$ phase diagram of Cu$_{3}$Mo$_{2}$O$_{9}$.
The shape of the symbols distinguishes 
the physical quantities to be used to obtain the phase boundary.
The triangles, squares, and circles denote 
the dielectric constants along the $a$- and $c$-axes and 
specific heat, respectively.
The solid (open) symbols denote 
the phase boundary 
obtained from the data of the $T$ ($H$) dependence.
}
\label{PhaseDiagramC}
\end{figure}
\begin{figure}
\begin{center}
\includegraphics[width=\textwidth,clip]{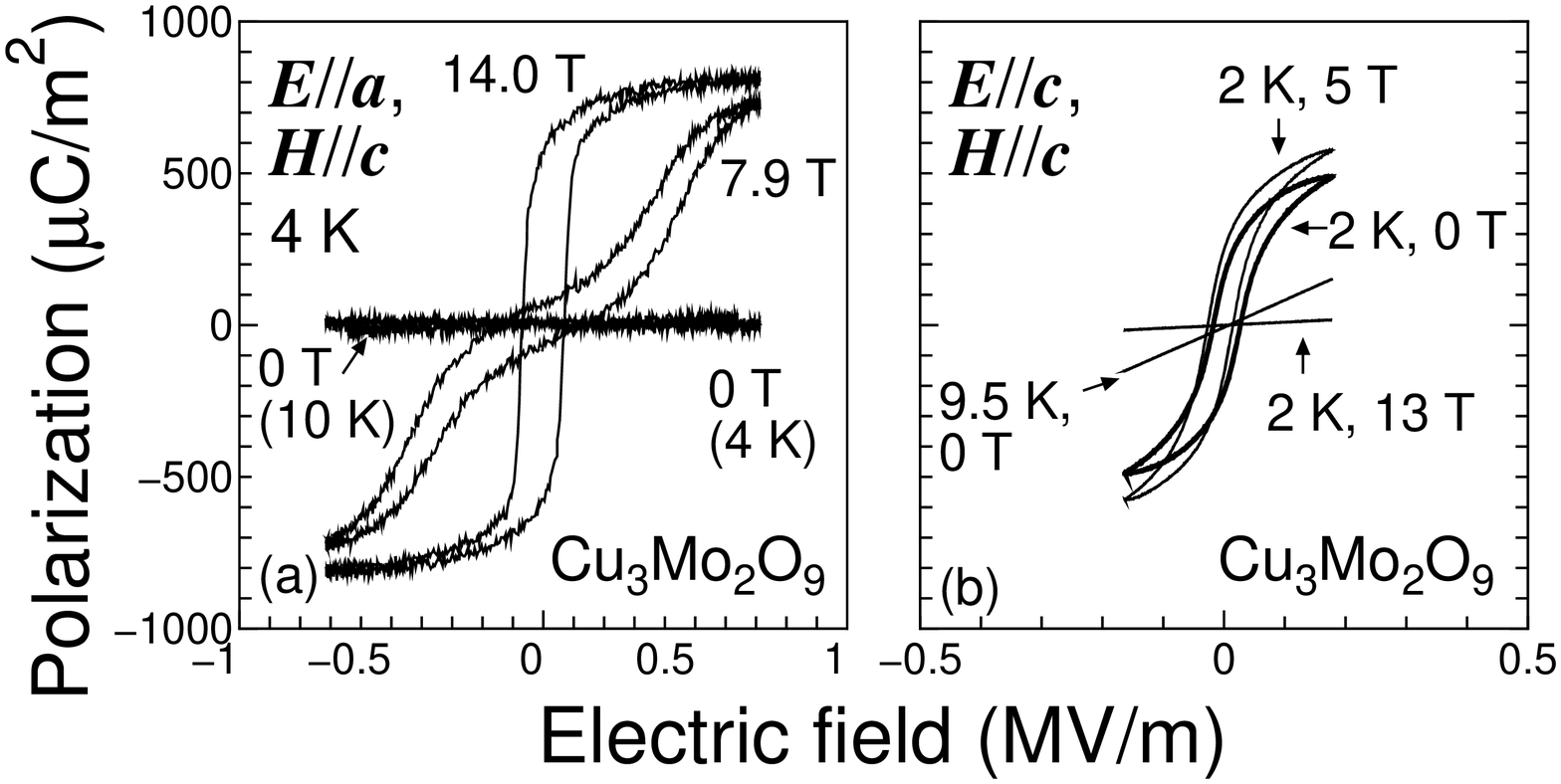}
\end{center}
\caption{
Typical polarization-electric-field loops.
The two loops at 4 and 10 K under 0 T in (a) 
almost overlap with each other. 
}
\label{PolarizationAC}
\end{figure}
Figures \ref{DielectricConstantC}(a) and \ref{DielectricConstantC}(b) show 
the typical $T$ dependences of $\epsilon_{\alpha}$ 
($\alpha$ = $a$ or $c$) 
under a fixed $H$ 
($\epsilon_{\alpha}$-$T$ curves), respectively, 
each of which has a (local) maximum 
$\epsilon_{\alpha}^{\rm peak}$ at $T_{\alpha}^{\rm peak}$.
The values of $T_{\alpha}^{\rm peak}$ against $H$ are 
plotted in the $H$-$T$ phase diagram 
by the solid symbols in Fig. \ref{PhaseDiagramC}.
Figures \ref{DielectricConstantC}(c) and \ref{DielectricConstantC}(d) show 
the typical $H$ dependences of $\epsilon_{\alpha}$ 
at 8 and 4 K 
($\epsilon_{\alpha}$-$H$ curves), respectively.
At 8 K, as shown in Fig. \ref{DielectricConstantC}(c),
the $\epsilon_{a}$-$H$ and $\epsilon_{c}$-$H$ 
curves have one and two peaks, respectively.
These are plotted in the phase diagram 
by open symbols in Fig. \ref{PhaseDiagramC}.
We observed 
a colossal magnetocapacitance (800\%) at 8 K 
in Fig. \ref{DielectricConstantC}(c).
Here the maximum $\epsilon_{a}$ slightly above 10 T 
was compared with the minimum $\epsilon_{a}$ at approximately 6 T.
This magnetocapacitance is larger than 
the maximum one reported before.~\cite{Goto2004}
This is probably due to 
strong fluctuations around the tricritical point 
at ($H$, $T$) = (10 T, 8 K) in the $H$-$T$ phase diagram.
At 4 K, as shown in Fig. \ref{DielectricConstantC}(d), 
$\epsilon_{c}$ gradually decreases 
with increasing $H$ above 6 T.
At about 8 T, 
a small peak is observed 
in the $\epsilon_{c}$-$H$ curve, 
which is indicated by an arrow
and is plotted in Fig. \ref{PhaseDiagramC}
by an open symbol.
Around this $H$, $\epsilon_{a}$ shows a sudden increase.

At 4 K, as shown in the inset of Fig. \ref{DielectricConstantC}(d), 
the $\epsilon_{c}$-$H$ curve 
under a zero-magnetic-field cooling process 
rapidly decreases with increasing $H$ from 0 T.
This anomaly is not observed during the reduction of the magnetic field.
We plot $H$, 
where this hysteresis effect disappears, 
in Fig. \ref{PhaseDiagramC}.
Around this $H$, the $H$ dependence of 
the magnetization $M_{c}$ (the $M_{c}$-$H$ curve) along the $c$-axis 
shows a small jump with a magnetic-field hysteresis effect.~\cite{Hamasaki2008}
In some multiferroic materials, 
a change in electric polarization accompanied by 
a jump of the magnetization 
has been reported.~\cite{KimuraNature2003,Kimura2006}
These facts and the present result showing that 
the peaks in the $\epsilon_{\alpha}$-$T$ curves become sharp 
at 6 T [Figs. \ref{DielectricConstantC}(a) and \ref{DielectricConstantC}(b)] 
suggest that the zero-field ground state contains some fluctuations, 
which is consistent with 
our picture describing the canted antiferromagnetism 
at zero and finite magnetic fields.~\cite{Hamasaki2008,Hase2008}
In this picture, the canting direction of the spin moment 
contains randomness after a zero-field cooling process.
The jump of the magnetization in finite magnetic fields 
is understood as the alignment of the canting direction.
Once it occurs, it survives even at a zero magnetic field 
because of the internal magnetic field, 
indicating the possibility of 
the magnetic-field hysteresis effect in this system.

Together with the additional data 
and phase boundary 
obtained from the $T$ dependence of 
the specific heat under $H$,~\cite{Hamasaki2010}
we obtain the $H$-$T$ phase diagram in Fig. \ref{PhaseDiagramC}.
One can see that the increase in $T_{\alpha}^{\rm peak}$ at 6 T 
[Figs. \ref{DielectricConstantC}(a) and \ref{DielectricConstantC}(b)]
corresponds to 
the change in $T_{\rm N}$ under $H$.~\cite{Hamasaki2010}
We emphasize here that the peak 
in the $\epsilon_{\alpha}$-$T$ ($\epsilon_{\alpha}$-$H$) 
curve is not necessarily in agreement with the critical temperature 
(magnetic field) of the phase transition.
Therefore, 
it is natural that 
the plots in the $H$-$T$ phase diagram 
obtained from the data of $\epsilon_{\alpha}$ 
do not perfectly trace the phase boundary 
obtained using specific heat.
We conclude that the phase diagram contains four different phases.

\medskip
To observe the spontaneous electric polarization directly, 
we measured the $P_{\alpha}$-$E_{\alpha}$ loop.
Figures \ref{PolarizationAC}(a) and \ref{PolarizationAC}(b) show 
the typical results at various $T$ and $H$.
At 0 T, 
as shown in Fig. \ref{PolarizationAC}(b), 
the $P_{c}$-$E_{c}$ loop below $T_{\rm N}$ 
shows clear ferroelectric behavior.
The spontaneous electric polarization density 
($\sim$500 $\mu$C/m$^{2}$) is 
close to that of TbMnO$_{3}$.~\cite{KimuraNature2003}
As shown in Fig. \ref{PolarizationAC}(a), 
the $P_{a}$-$E_{a}$ loop 
at 0 T does not show ferroelectric behavior.
Thus, 
we conclude that Cu$_{3}$Mo$_{2}$O$_{9}$ 
at 0 T below $T_{\rm N}$ is 
in the ferroelectric phase with spontaneous electric polarization 
along the $c$-axis [the FE(${\boldsymbol P}_{s}//{\boldsymbol c})$ phase].
This result is consistent with the following facts:
$\epsilon_{c}^{\rm peak} \sim 10 \times \epsilon_{a}^{\rm peak}$ 
in Figs. \ref{DielectricConstantC}(a) and \ref{DielectricConstantC}(b) and 
a strong microwave absorption was observed 
at approximately $T_{\rm N}$ at a zero magnetic field.

Figure \ref{PolarizationAC}(b) shows 
the $H$ dependence of the $P_{c}$-$E_{c}$ loop.
At 5 T, the $P_{c}$-$E_{c}$ loop 
indicates the FE(${\boldsymbol P}_{s}//{\boldsymbol c})$ phase
because it is similar to that at 0 T (data not shown).
Instead of the closed $P_{c}$-$E_{c}$ loop at 13 T, 
the ferroelectric $P_{a}$-$E_{a}$ loop was observed at 14 T.
The $\epsilon_{a}^{\rm peak}$ at 13 T becomes about ten times larger 
than $\epsilon_{a}^{\rm peak}$ at 6 T [Fig. \ref{DielectricConstantC}(a)].
The $\epsilon_{c}^{\rm peak}$ at 13 T becomes about ten times smaller 
than $\epsilon_{c}^{\rm peak}$ at 6 T [Fig. \ref{DielectricConstantC}(b)].
Judging from these results, 
we confirm the FE(${\boldsymbol P}_{s}//{\boldsymbol a})$ phase at 13 T.
The spontaneous electric polarization density 
($\sim$800 $\mu$C/m$^{2}$) is 
comparable to that of DyMnO$_{3}$ at 21 K.~\cite{Goto2004}

As shown in Fig. \ref{DielectricConstantC}(d), 
both the $\epsilon_{a}$-$H$ and $\epsilon_{c}$-$H$ curves 
show anomalies at approximately 8 T at 4 K.
As shown in Fig. \ref{PolarizationAC}(a), 
the $P_{a}$-$E_{a}$ curve suggests 
a double hysteresis loop at 7.9 T.
This effect is explained by 
the electric-field-induced electric polarization 
and indicates strong fluctuations around the phase boundary.
We conclude that 
the change in the direction of the spontaneous electric polarization
occurs at the phase boundary 
running from ($H$, $T$) = (8 T, 2 K) to (10 T, 8 K).
At approximately 8 T, 
a change in the electron-spin-resonance spectrum 
was reported.~\cite{Okubo2010}
Together with 
the phase transition temperatures 
obtained from the $T$ dependences of $M_{c}$ 
under various $H$,~\cite{Hamasaki2010}
we conclude that 
this compound is a multiferroic material.

\medskip
In many cases, 
the origin of multiferroicity has been discussed 
on the basis of the formation of the magnetic superlattice.
Unfortunately, 
the spin structure below $T_{\rm N}$ at 0 T 
has not been clarified yet. 
However, we emphasize again that 
the AFM long-range order below $T_{\rm N}$ 
has been established 
by the magnetic dispersion curve obtained 
from inelastic neutron scattering.~\cite{KuroeSubmittedToPRB}
Moreover, 
the anisotropic magnetization 
has been quantitatively explained on the basis of a weakly canted 
AFM long-range order.~\cite{Hamasaki2008}

\medskip
\begin{figure}
\begin{center}
\includegraphics[width=0.7\textwidth]{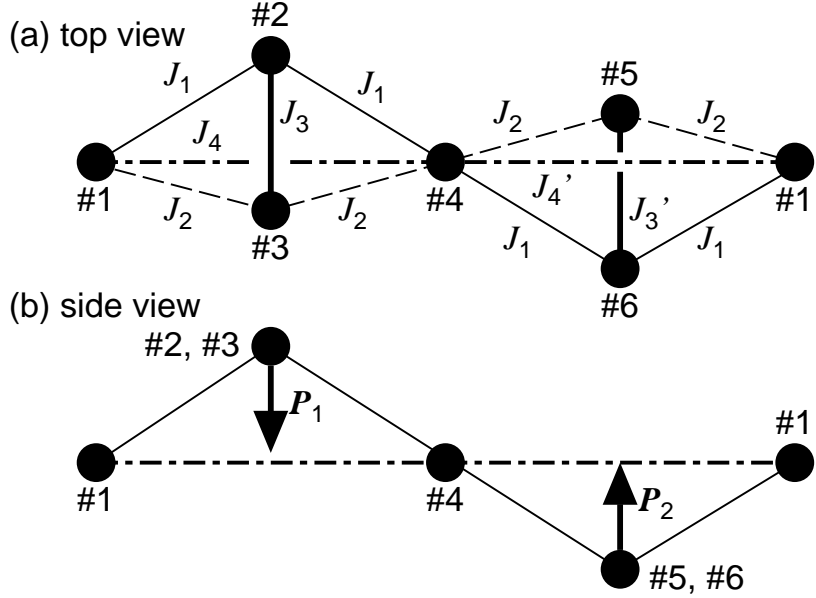}
\end{center}
\caption{Six-spin cluster under periodic-boundary condition.
\#$i$ indicates the $i$th $S$=1/2 spin.
This spin cluster can be divided into 
two tetrahedral spin clusters, 
each of which has an electric polarization 
${\boldsymbol P}_{1}$ or ${\boldsymbol P}_{2}$.
The charge redistribution at the center site (\#4) 
is divided so that both 
${\boldsymbol P}_{1}$ and ${\boldsymbol P}_{2}$ 
exist on the bisector between sites \#1 and \#4.
}
\label{SpinSystems2}
\end{figure}
The point of discussion is 
how the AFM long-range order induces ferroelectricity.
In the following, 
we focus on the charge redistribution effects 
caused by the three-spin ring exchange interaction 
in geometrically frustrated Mott insulators, 
proposed by Bulaevskii {\it et al.}~\cite{Bulaevskii2008,Khomskii2010}
We demonstrate that 
the spin cluster in Fig. \ref{SpinSystems2}
has the potential for multiferroic behavior
as a result of 
the {\em antiferroelectric} (AFE) polarization 
at zero electric field and 
the {\em ferrielectric} (FRE) polarization 
under an electric field.
This is the minimum system 
that maintains the tetrahedral spin arrangement and 
the inversion center at spin site \#4 simultaneously.
The formation of the magnetic superlattice is impossible.

We treat this system as a Mott insulator.
The electronic band from 
$d_{x^2-y^2}$ orbitals in Cu$^{2+}$ ions is half-filled.
In the limit of the strong on-site Coulomb repulsion $U$, 
where the charge degrees of freedom are frozen, 
the system can be mapped on the quantum spin system.
The spin Hamiltonian is given by 
\begin{equation}
{\cal H} = \sum_{\langle i,j \rangle}\frac{4t_{ij}^{2}}{U}
{\boldsymbol S}_{i} \! \cdot \! {\boldsymbol S}_{j} 
+ \sum_{i} g \mu_{\rm B} 
{\boldsymbol S}_{i} \! \cdot \! 
{\boldsymbol H}_{i}^{\rm loc} 
\ ,
\end{equation}
where the sum in the first term on the right-hand side runs over 
all the possible spin pairs (${\boldsymbol S}_{i}$ and 
${\boldsymbol S}_{j}$ at the \#$i$ and \#$j$ sites, 
respectively) 
connected through the hopping parameter $t_{ij}$.
The values of $4t_{ij}^2/U$ correspond 
to the exchange interactions 
in Fig. \ref{SpinSystems2}.
The second term on the right-hand side is 
the magnetic energy from the local magnetic field 
${\boldsymbol H}^{\rm loc}_{i}$ at the \#$i$ site, 
where $g$ and $\mu_{\rm B}$ are the $g$ factor 
and Bohr magneton, respectively.

On the basis of the interchain interaction 
in Cu$_{3}$Mo$_{2}$O$_{9}$,~\cite{KuroeSubmittedToPRB}
we set the values $J_{1}$ = $J_{2}$ = 1 meV, $J_{3}$ = $J_{3}'$ = 5.8 meV, 
$J_{4}$ = 4 meV, and $g$ = 2. 
We introduced the staggered magnetic field 
$H_{\rm AFM}$ $\equiv$ $|{\boldsymbol H}^{\rm loc}_{i}|$
along the quantization axis working on only the \#1 and \#4 sites.
$H_{\rm AFM}$ induces the AFM spin order 
at the site on the spatial inversion symmetry, 
i.e., this site is on the anti-inversion center 
in the term of the magnetic space group.
Using the exact diagonalization, 
we calculated 
$\langle {\boldsymbol S}_{i} \rangle$ and 
the charge redistribution $\delta n_{i}$ 
at spin site \#$i$,~\cite{Bulaevskii2008,Khomskii2010}
\begin{equation}
\delta n_{i} \! = \! \sum_{\langle i,j,k \rangle}
\frac{8t_{ij}t_{jk}t_{ki}}{U^3}
\left[
{\boldsymbol S}_{i} \! \cdot \! 
\left({\boldsymbol S}_{j} + {\boldsymbol S}_{k} \right)
-2 {\boldsymbol S}_{j} \! \cdot \! {\boldsymbol S}_{k} 
\right]
\ ,
\end{equation}
in the periodic-boundary six-spin cluster 
at 0 K.
Here, the sum on the right-hand side runs over 
all the possible spin triangles 
connected through exchange interactions.
The amplitude of the electric polarization 
$|{\boldsymbol P}_{1(2)}|$ is proportional 
to $\delta n_{1(5)} - \delta n_{2(1)} - \delta n_{3(4)} +\delta n_{4(6)}$, 

\begin{figure}
\begin{center}
\includegraphics[width=\textwidth]{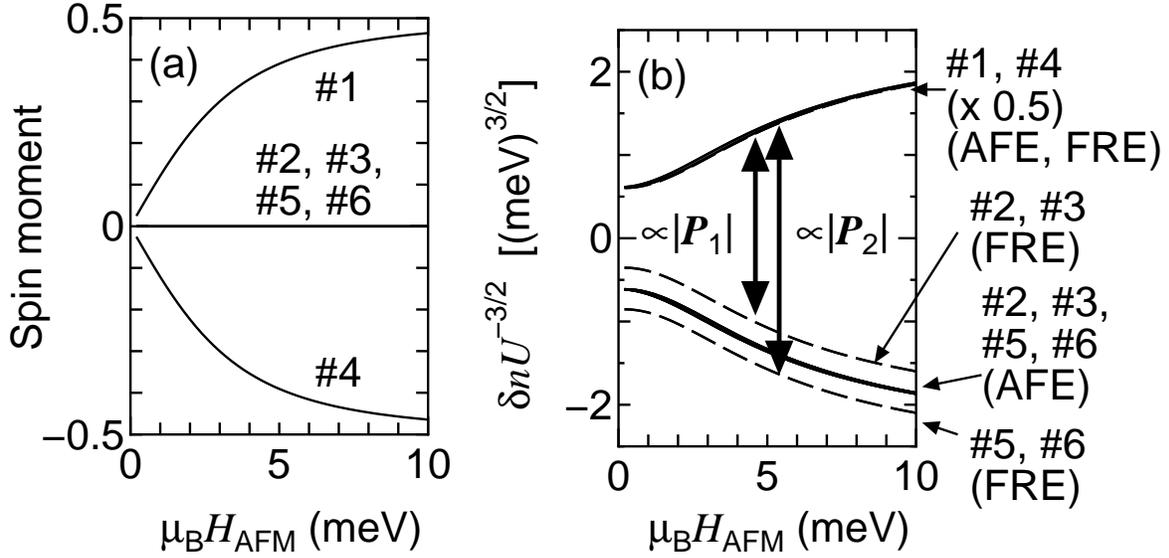}
\end{center}
\caption{Spin moments projected on quantum axis (a) 
and charge-redistribution parameters (b) 
in six-spin cluster under periodic-boundary condition 
as functions of staggered magnetic field.
The parameter sets that give the antiferroelectric (AFE) 
and ferrielectric (FRE) phases are given in the text.
The electric polarizations are shown by the arrows.
}
\label{Calculation6sites}
\end{figure}
As shown in Figs. \ref{Calculation6sites}(a) and \ref{Calculation6sites}(b), 
$|\langle {\boldsymbol S}_{\{1,4\}} \rangle_{z}|$ 
and $|{\boldsymbol P_{\{1,2\}}}|$
increase larger with increasing 
$H_{\rm AFM}$.
Even in this case, 
the spins at the sites \#2, \#3, \#5, and \#6 
still form nonmagnetic spin dimers.
We plot $\delta n U^{-3/2}$ 
in Fig. \ref{Calculation6sites}(b)
because we could not precisely estimate $U$, 
and calculations in larger systems are necessary
to discuss the origin of 
the finite $|{\boldsymbol P}_{\{1,2\}}|$ 
without the staggered magnetic field 
because of the strong system-size dependence.

Under a uniform electric field, 
the exchange interaction slightly changes 
as a result of the breaking of 
the spatial inversion symmetry.~\cite{Triff2010}
If the exchange interactions 
of the nonmagnetic spin dimers are different
($J_{3}$ = 5.0 meV and $J_{3}'$ = 6.6 meV), 
we obtain a finite net electric dipole moment 
as a result of the FRE alignment of the electric dipoles 
(${\boldsymbol P}_{1}$ $\neq$ $-{\boldsymbol P}_{2}$)
even though $\langle {\boldsymbol S}_{i} \rangle$ is not changed.
From this calculation, we conclude that 
the tetrahedral quasi-one-dimensional spin system has 
the potential to exhibit AFM-AFE and AFM-FRE type multiferroic behaviors.
The finite net electric dipole moment appears 
when the time and spatial inversion symmetries are broken.

\medskip
In this study, 
we demonstrate the possibility of 
electric polarization induced by 
the AFM long-range order
in a distorted tetrahedral quasi-one-dimensional spin system.
As experimental evidence, 
the $H$-$T$ phase diagram of Cu$_{3}$Mo$_{2}$O$_{9}$ 
under a magnetic field along the $c$-axis was shown 
by studying the temperature and magnetic-field
dependences of the dielectric constant.
The ferroelectric behavior was observed 
in the polarization-electric-field loop and 
a change in the polarization direction 
was observed.
Around the tricritical point at (10 T, 8 K), 
a colossal magnetocapacitance 
was observed.
We showed that 
a six-spin cluster that corresponds to 
a tetragonal quasi-one-dimensional spin system 
has the potential to become an 
AFM-AFE multiferroic state 
without the formation of a magnetic superlattice.
The possibility of the AFM-FRE multiferroic state 
under an electric field was discussed.


\medskip
\section*{Acknowledgments}
This work is partly supported by 
a Grants-in-Aid for Scientific Research (C) (No. 40296885) 
and on Priority Area (No. 19052005) 
from the Ministry of Education, Culture, Sports, 
Science and Technology of Japan (MEXT).
We thank Dr. N. Terada of NIMS, 
Dr. Y. Nishiwaki and Prof. T. Nakamura 
of Shibaura Institute of Technology, and 
Prof. T. Goto, Dr. M. Akaki, and 
Prof. H. Kuwahara of Sophia University 
for helpful discussions.
We also wish to acknowledge 
the technical assistance of Mr. R. Kino and Mr. M. Suzuki.

\end{document}